%% file: prep.tex
\documentstyle[aaspp4]{article}
\slugcomment{ApJ, in press}

\begin{document}

\title{Effects of Ram-Pressure from Intracluster Medium on the 
  Star Formation 
  Rate of Disk Galaxies in Clusters of Galaxies}
\author{Yutaka Fujita and Masahiro Nagashima}
\affil{Department of 
  Earth and Space Science,
  Graduate School of Science,
  Osaka University,
  Machikaneyama-cho, 
  Toyonaka, Osaka, 560-0043, Japan}
\authoremail{fujita@vega.ess.sci.osaka-u.ac.jp}

\begin{abstract}

  Using a simple model of molecular cloud evolution, we have
  quantitatively estimated the change of star formation rate (SFR) of
  a disk galaxy falling radially into the potential well of a cluster
  of galaxies. The SFR is affected by the ram-pressure from the
  intracluster medium (ICM).  As the galaxy approaches the cluster
  center, the SFR increases to twice the initial value, at most, in a
  cluster with high gas density and deep potential well, or with a
  central pressure of $\sim 10^{-2}\rm\: cm^{-3}\; keV$
  because the ram-pressure compresses the molecular gas of the galaxy.
  However, this increase does not affect the color of the galaxy
  significantly. Further into the central region of the cluster
  ($\lesssim 1$ Mpc from the center), the SFR of the disk component
  drops rapidly due to the effect of ram-pressure stripping. This
  makes the color of the galaxy redder and makes the disk dark. These
  effects may explain the observed color, morphology distribution and
  evolution of galaxies in high-redshift clusters. By contrast, in a
  cluster with low gas density and shallow potential well, or the
  central pressure of $\sim 10^{-3}\rm\: cm^{-3}\; keV$, the SFR of a
  radially infalling galaxy changes less significantly, because
  neither ram-pressure compression nor stripping is effective.
  Therefore, the color of galaxies in poor clusters is as blue as that
  of field galaxies, if other environmental effects such as
  galaxy-galaxy interaction are not effective. The predictions of the
  model are compared with observations.

\end{abstract}


\section{Introduction}
\label{sec:intro}

Clusters of galaxies in the redshift range of $0.2-0.5$ often exhibit
an overabundance, relative to present-day clusters, of blue galaxies
(Butcher \& Oemler \markcite{bo1978}1978). This star formation
activity is often called the Butcher-Oemler effect (BOE).  Although
several mechanisms have been suggested to explain the effect, it
is still not known which is most important. This is because the
change of star formation rate (SFR) caused by these mechanisms has not
been predicted physically and quantitatively, which makes the
comparison between the theoretical predictions and observations
difficult.

In contrast with the usual ``black-box'' approach of N-body
simulations which has previously dominated the theoretical discussion
in this field, Fujita \markcite{f1998}(1998, Paper I) constructs a
simple model to evaluate the change of SFR of a disk galaxy made by
environmental effects in clusters of galaxies. He shows that tidal
force from the potential well of a cluster can induce star formation
activity of a galaxy in the central part of the cluster, but it is
inconsistent with observations. He also shows that the increase of
external thermal pressure when a galaxy plows into the intracluster
medium (ICM) does not induce significant rise of the SFR. On the
contrary, he shows that successive high-speed encounters between
galaxies (galaxy harassment, see Moore et al. 1996\markcite{mkl1996})
can make star formation active.

These arguments are based on the idea that the BOE is caused by blue
starburst galaxies. Some recent observations, however, suggest that
starbursts do not play a major role in driving cluster galaxy
evolution at high redshift. Abraham et al. \markcite{ash1996}(1996)
observe Abell 2390 and find that the galaxy population in the cluster
changes gradually from a red, evolved, early-type population in the
inner part of the cluster to a progressively blue, later-type
population in the extensive outer envelope of the cluster. The radial
change of colors and morphologies of galaxies is also observed in
other clusters (e.g. Balogh et al. \markcite{bmy1997}1997; Rakos,
Odell, \& Schombert \markcite{ros1997}1997 ; Oemler, Dressler, \&
Butcher \markcite{odb1997}1997 ; Smail et al. \markcite{see1998}1998 ;
Couch et al. \markcite{cbs1998}1998 ; van Dokkum et
al. \markcite{vfk1998}1998).  Abraham et al. \markcite{ash1996}(1996)
speculate that the cluster has been built up gradually by the infall
of blue field galaxies and that star formation has been truncated in
infalling galaxies during the accretion process. The BOE is due to a
high proportion of blue spiral galaxies which later turn to red S0
galaxies during the infall into the cluster center. Starbursts are not
required to explain the galaxy spectra.  In fact, high angular
resolution images from the {\em Hubble Space Telescope (HST)} show
that significant fraction of blue galaxies in high redshift clusters
are ``normal'' spiral galaxies (Oemler et al. \markcite{odb1997}1997).
Moreover, using {\em HST} data, Couch et al. \markcite{cbs1998}(1998)
claim that there are fewer disturbed galaxies than predicted by the
galaxy harassment model. They suggest that ram-pressure stripping by
hot ICM truncates star formation of blue disk galaxies and makes them
red.

Motivated by these observations, the effects of ram-pressure from ICM
on the SFR of a disk galaxy are investigated in this study.  Previous
theoretical studies show that in the central part of clusters, where
the density of the ICM is large, ram-pressure from the ICM sweeps
interstellar medium (ISM) of galaxies away (e.g. Gunn \& Gott
\markcite{gg1972}1972; Takeda, Nulsen, \& Fabian
\markcite{tnf1984}1984; Gaetz, Salpeter, \& Shaviv
\markcite{gss1987}1987). Thus, the SFR could decrease as Couch et
al. \markcite{cbs1998}(1998) suggest.  On the other hand, ram-pressure
could enhance star formation through the compression of ISM. For
example, Bothun \& Dressler \markcite{bd1986}(1986) suggest that the
observational data of 3C295 cluster and Coma cluster are consistent
with the picture of ram-pressure induced star formation.  Therefore,
the quantitative estimation of the SFR of a galaxy under pressure from
the ICM is required to know whether the ram-pressure decreases or
increases the SFR.  The plan of this paper is as follows. In
\S\ref{sec:model}, the models of molecular gas and ram-pressure are
described. In \S\ref{sec:evol}, the evolution of the SFR of a
infalling galaxy is investigated. In \S\ref{sec:dis}, the results of
the model calculations are compared with observations. The conclusions
are summarized in \S\ref{sec:conc}.

\section{Models}
\label{sec:model}

\subsection{Molecular Clouds}
\label{sec:mol}

The method for modeling the evolution of the molecular clouds in disk
galaxies was extensively discussed in Paper I. In this subsection the
main features of the method are briefly reviewed. 
Molecular clouds are divided according to their {\em initial} 
masses, that is, $M_{\rm
  min}=M_1< ...< M_i< ...<M_{\rm max}$, where the
lower and upper cutoffs are $M_{\rm
  min}=10^2\;\rm M_{\sun}$ and
$M_{\rm max}=10^{8.5}\;\rm M_{\sun}$, respectively, and their intervals
are chosen so that $\log(M_{i+1}/M_i)= 0.01$.  
Referring to the total mass of clouds whose initial masses are
between $M_i$ and $M_{i+1}$ as $\Delta \tilde{M}_i(t)$, the rate of 
change is
\begin{equation}
  \label{eq:rate}
  \frac{d \Delta \tilde{M}_i(t)}{dt} 
  = \tilde{f}_i S(t) - \frac{\Delta 
    \tilde{M}_i(t)}{\tau(M_i,P)} \;,
\end{equation}
\begin{equation}
  \label{eq:sor}
  S(t) = S_{\star}(t)+S_{\rm mol}(t) \;,
\end{equation}
where $\tilde{f}_i$ is the initial mass fraction of the molecular clouds 
whose initial masses are
between $M_i$ and $M_{i+1}$, 
$S_{\star}(t)$ is the gas ejection rate of stars, $S_{\rm mol}(t)$ is
the recycle rate of molecular gas, 
$\tau(M_i,P)$ is the 
destruction time of a molecular cloud with mass $M_i$ and pressure $P$.
The initial mass function of molecular clouds is
assumed to be $N(M) \propto M^{-2}$. Thus $\tilde{f}_i$ can be derived 
from this relation.

The gas ejection rate of stars is divided into two terms, 
$S_{\star}(t) = S_s(t) + S_l$.
The gas ejection rate of stars with small lifetime is given by
\begin{equation}
  \label{eq:sou}
  S_s(t) = \int^{m_u}_{m_l} \psi(t-t_m)R(m)\phi(m)dm \:,
\end{equation}
where $m$ is the stellar mass, $\psi(t)$ is the SFR, $\phi(m)$ is the
IMF expressed in the form of the mass fraction, $R(m)$ is the return
mass fraction, and $t_m$ is the lifetime of stars with mass $m$.  The
slope of the IMF is taken to be 1.35 (Salpeter mass function). The
upper and lower mass limits, $m_u$ and $m_l$, are taken to be $50
M_{\sun}$ and $2.1 M_{\sun}$, respectively, although the lower limit
of the IMF is $0.08 M_{\sun}$. The gas ejection rate of stars with
mass smaller than $m_l$, or $S_l$, is assumed to be constant. It is
determined by the balance between the formation and destruction rate
of molecular clouds (see Paper I).

The recycle rate of molecular gas is 
\begin{equation}
  \label{eq:rec}
  S_{\rm mol}(t) = \sum_i [1-\epsilon(M_i,P)]\frac{\Delta 
    \tilde{M}_i(t)}{\tau(M_i,P)}\;,
\end{equation}
where $\epsilon(M_i,P)$ is the star formation efficiency 
of a molecular cloud with mass $M_i$ and pressure $P$.
Note that $S_{\rm mol}(t)$ is identical with the
evaporation rate of molecular gas by newborn stars. 
On simple assumptions, Elmegreen \& Efremov
\markcite{ee1997}(1997) derive $\epsilon(M_i,P)$ and
$\tau(M_i,P)$ ; their results are adopted hereafter. 

The SFR is described by  
\begin{equation}
  \label{eq:sf}
  \psi(t) = \sum_i \epsilon(M_i,P)\frac{\Delta 
    \tilde{M}_i(t)}{\tau(M_i,P)}\;.
\end{equation}

Equation (\ref{eq:rate}) can be solved if $P$ is given. 
In the next subsection, $P$ is determined for ram-pressure. Then the 
SFR is calculated from equations (\ref{eq:rate})
$-$ (\ref{eq:sf}).

\subsection{Ram-pressure}
\label{sec:ram}

As mentioned in \S\ref{sec:intro}, ram-pressure could induce more star
formation in a disk galaxy through compression of the ISM,
while it could decrease the SFR by stripping the ISM from a
galaxy.  In this subsection a model is constructed to determine which 
effect dominates.

In addition to molecular gas, warm HI gas is also considered, although
equation (\ref{eq:rate}) does not include it explicitly.  It is
assumed that the gas ejected from stars and the gas evaporated by
young stars temporary become HI gas before they finally become
molecular gas.  The time-scale of the HI gas phase is assumed to be
small in comparison with that of the galaxy evolution.  Thus,
the total mass of HI gas can be given by $M_{\rm HI}
\approx \tau_{\rm mol}S/\epsilon_{\rm mol}$, where $\tau_{\rm mol}$
and $\epsilon_{\rm mol}$ are the time-scale and efficiency of
molecular cloud formation, respectively.  If molecular cloud formation
is triggered by density waves in the galactic disk, $\tau_{\rm mol}$ is
nearly the ratio of the rotation time of the galaxy to the number of
the arms.  In this case, it can be assumed that $\tau_{\rm mol}$ is
constant. If $\epsilon_{\rm mol}$ is also constant, $M_{\rm HI}
\propto S(t)$. Note that $S(t)$ changes only $\lesssim 20$ \% in the
calculations in \S\ref{sec:evol}.

Ram-pressure from ICM is
\begin{equation}
\label{eq:ram}
P_{\rm ram} = \rho_{\rm ICM} v_{\rm gal}^2 \:,
\end{equation} 
where $\rho_{\rm ICM}$ is the mass density of ICM, and $v_{\rm gal}$
is the velocity of a galaxy relative to ICM. The ram-pressure is
assumed to be the same both for molecular gas and for HI gas.  For the
HI gas, the restoring force per unit area due to the gravity of the
disk of the galaxy is given by
\begin{eqnarray}
  F_{\rm grav,HI}
  &=& 2\pi G \Sigma_{\star} \Sigma_{\rm HI} \\
  &=& v_{\rm rot}^2 R^{-1} \Sigma_{\rm HI} \label{eq:grav2} \\
              &=& 2.1\times 10^{-11}{\rm dyne\: cm^{-2}}
               \left(\frac{v_{\rm rot}}{220\rm\; km\: s^{-1}}\right)^2
               \left(\frac{R}{10\rm\; kpc}\right)^{-1}
               \left(\frac{\Sigma_{\rm HI}}{8\times 10^{20} 
                   m_{\rm H}\;\rm cm^ {-2}}\right) \label{eq:grav3}\:, 
\end{eqnarray}
where $G$ is the gravitational constant, $\Sigma_{\star}$ is the
gravitational surface mass density, $\Sigma_{\rm HI}$ is the surface
density of the HI gas, $v_{\rm rot}$ is the rotation velocity of the
galaxy, $R$ is the radius of the disk, and $m_{\rm H}$ is the mass of
hydrogen (Gunn \& Gott \markcite{gg1972}1972). Equation
(\ref{eq:grav2}) is derived from the relation $\Sigma_{\star}=(2\pi
G)^{-1} v_{\rm rot}^2 R^{-1}$ (Binney \& Tremaine
1987\markcite{bt1987}). In the following section, $v_{\rm
  rot}=220\;\rm km\; s^{-1}$ and $R=10$ kpc are used unless otherwise
mentioned. Assuming $\Sigma_{\rm HI} \propto M_{\rm HI}$, the surface
density can be given by $\Sigma_{\rm HI} = \Sigma_{\rm
  HI,0}(S(t)/S_0)$, where $\Sigma_{\rm HI,0}$ and $S_0$ are the
initial surface density and SFR, respectively.

The HI gas is stripped when
\begin{equation}
  \label{eq:strip}
 P_{\rm ram}>F_{\rm grav,HI} \:. 
\end{equation}
After the HI gas is stripped, instead of equation (\ref{eq:sor}),
$S(t)$ is fixed at zero in equation (\ref{eq:rate}).

Molecular clouds are also stripped when 
\begin{equation}
  \label{eq:str_mol}
  P_{\rm ram}>2\pi G \Sigma_{\star} \Sigma_{\rm mol} \:,
\end{equation}
where $\Sigma_{\rm mol}$ is the
column density of each molecular cloud and is given by
\begin{equation}
  \label{eq:sig_mol}
  \Sigma_{\rm mol} = 190\; {\rm M_{\sun}pc^{-2}}
         \left(\frac{P_{\rm ram}}{10^4\; k_{\rm B}\rm\; cm^{-3}\; K}
         \right)^{1/2} \:,
\end{equation}
regardless of the mass 
(Elmegreen \markcite{e1989}1989). Since $\Sigma_{\rm mol}>
\;\Sigma_{\rm HI}$, the molecular clouds
are stripped after the HI gas is stripped.   

\section{Evolution of Radially Infalling Galaxy}
\label{sec:evol}

The distribution of the gravitational matter of a model cluster is
\begin{equation}
  \label{eq:grav_cl}
  \rho(r) = \frac{\rho_0}{(1+r^2/r_c^2)^{3/2}} \:, 
\end{equation}
where $r$ is the distance from the cluster center, $\rho_0$ is the
mass density at the center, and $r_c \;(=400\;\rm kpc)$ is the core
radius of the cluster.  The central mass density is given by
\begin{eqnarray}
  \label{eq:rho0}
  \rho_0 &=& \frac{9}{4\pi G r_c^2}\frac{k_{\rm B}T}{\mu m_{\rm H}} \\
         &=& 9.0\times 10^{-26} 
             \;{\rm g\; cm^{-3}}\left(\frac{r_c}{400\;\rm
             kpc}\right)^{-2} 
             \left(\frac{k_{\rm B}T}{8\;\rm keV}\right) \:,
\end{eqnarray}
where $k_{\rm B}$ is the Boltzmann constant, $T$ is the effective
temperature of the cluster, and $\mu (=0.6)$ is the mean molecular
weight.  Note that recent numerical simulations support a steeper core
profile (e.g. Navarro, Frenk, \& White
\markcite{nfw1997}1997). However, even if this profile is adopted
instead of equation (\ref{eq:grav_cl}), the following results does not
change significantly (see Figure 2 in Paper I).  

The gas distribution is
\begin{equation}
  \label{eq:gas_cl}
  \rho_{\rm gas}(r) = \frac{\rho_{\rm gas,0}}{1+r^2/r_c^2} \:,
\end{equation}
where $\rho_{\rm gas,0}$ is the central gas density.
The pressure of molecular clouds is given by
\begin{equation}
  \label{eq:press}
  P = \max(P_{\rm ram}+P_{\rm stat}, P_{\sun}) \:,
\end{equation}
where $P_{\sun}=3\times 10^4\;\rm cm^{-3}\; K$ is the pressure from
random motion of the clouds in the galaxy and $P_{\rm stat}=\rho_{\rm
  gas}k_{\rm B}T/\mu m_{\rm H}$ is the static pressure from ICM.

The initial surface density of HI gas and the initial SFR of the model
galaxy are $\Sigma_{\rm HI,0}=8\times 10^{20}m_{\rm H}\rm\; cm^{-2}$
and $S_0= 6\;\rm M_{\sun}\; yr^{-1}$, respectively. 
The parameters for the gas distribution of model clusters
are summarized in Table 1. They are the typical ones obtained by X-ray
satellites (e.g. Jones \& Forman 1984\markcite{jf1984} ; White, Jones, \&
Forman \markcite{wjf1997}1997). The initial position, $r_0$, and
velocity, $v_{\rm gal,0}$, of the model galaxy mean that the galaxy
begins to fall from where the mass density of the cluster is 100 times
the critical density of the Universe (for $H_0=75 \rm km\; s^{-1}\;
Mpc^{-1}$).  Moreover, if $P(t=0)$
is given, the initial total
mass of molecular clouds, $M_{\rm mol,0}$, and the gas
ejection rate of stars with long lifetime, $S_l$, can be determined so
that the formation and destruction rate of molecular clouds are
balanced at $t=0$ (see Paper I).  
For the parameters above, $M_{\rm mol,0}=2.5\times
10^9\;\rm M_{\sun}$ (except for models A1 and B1), $M_{\rm
  mol,0}=2.1\times 10^9\;\rm M_{\sun}$ (models A1 and B1), and $S_l
\approx 5\;\rm M_{\sun}\; yr^{-1}$. In models B1--B4, the stripping of
HI gas and molecular gas are ignored for comparison.

Figure 1 shows the pressure evolution. As is shown, they strongly
depend on the gas density and effective temperature of the cluster
because of equation (\ref{eq:ram}) and the dependence of galaxy
velocity on the depth of potential well. Figure 2 shows the SFR of a
model galaxy. The calculations continue even after the galaxy passes
the cluster center. The SFR rises to twice its initial value at
most. The maximum of the SFR can be explained by the relation between
pressure and destruction time of molecular clouds (Figure 4 in
Elmegreen \& Efremov \markcite{ee1997}1997), which is used in equation
(\ref{eq:rate}).  Model B2 is taken as an example. As a galaxy gets
closer to the cluster's center, the ram-pressure increases. Thus, the
destruction time of molecular clouds decreases. When $P\sim 10\;
P_{\sun}$ (at $r\sim 400$ kpc), the destruction times of all molecular
clouds ($<10^{8.5}\;\rm M_{\sun}$) become less than the time passed
since $P$ started to rise significantly ($\sim 2\times 10^8$ yr, see
Figure 1a and 4). Therefore, the pressure increase has affected all
the clouds by this time. In fact, the ratio of the initial total mass
of molecular clouds, $2.5\times 10^9\;\rm M_{\sun}$ to the time,
$2\times 10^8$ yr, corresponds to the peak of SFR, $\sim 13\;\rm
M_{\sun}\; yr^{-1}$.  Note that the total mass of molecular clouds at
$r=400$ kpc is not zero (Figure 3a). This is because the stars with
long lifetime supply additional gas.

The SFR rapidly drops in the central region of a high-temperature
and/or gas-rich cluster when ram-pressure stripping is considered
(models A1-A3).  This is because HI gas is stripped and the formation
of molecular clouds ceases. For example, the condition
(\ref{eq:strip}) is satisfied for $r < 1.2$ Mpc in model A1.  In that
region, without the supply ($S(t)=0$ in equation [\ref{eq:rate}]),
part of the molecular gas is consumed to form stars and the rest is
stripped soon after being evaporated by newborn stars. The larger the
density and temperature of the model cluster, the greater the effect
of ram-pressure, so that gas can be stripped and the SFR drops further
out in the cluster (Figure 1). If surface density of the galaxy is
smaller, the condition (\ref{eq:strip}) is satisfied at larger radius.
In that case, the SFR begins to decrease earlier.
In model A4, the condition (\ref{eq:strip}) is
never satisfied, that is, the HI gas is not stripped. In this model,
the SFR does not change significantly because neither ram-pressure
compression nor stripping is effective. Note that the ram-pressure
cannot strip molecular clouds in all the models considered here.

Although the SFR in models A1--A4 is that of all the galaxy, it is not
very different from that of the disk component because most gas
resides in the disk and because the parameter $R=10$ kpc is the
typical value of disk radius. The SFR of the bulge component can be
considered separately as follows. The normalized SFR, $\psi/S_0$, does
not depend on $S_0$, if the formation and destruction rate of
molecular clouds are initially balanced. Thus the normalized SFR in
models B1--B4 which do not include the effect of the ram-pressure
stripping corresponds to that of the galactic bulge where restoring
force is large (see equation [\ref{eq:grav3}]). For example, at $R=10$
kpc, the ram-pressure stripping occurs when $P\sim 10\; P_{\sun}$
(models A1--A3). Thus, for $R<0.5$ kpc, it occurs when $P\gtrsim 200\;
P_{\sun}$ if $v_{\rm rot}$ and $\Sigma_{\rm HI}$ are identical. That
is, the ram-pressure stripping does not occur in the bulge (Figure 1).
In models B1--B4, after $\psi/S_0$ rises only twice as much as the
initial value, it begins to decrease due to the shortage of molecular
gas (Figure 3). However, the star formation does not cease even if the
galaxy reaches the center of the cluster. After the galaxy passes the
center, the SFR continues to decrease, although the rate is slow
(Figure 2).

\section{Discussion}
\label{sec:dis}

In this section, the results in \S\ref{sec:evol} are compared with the
recent observations of distant clusters in order to investigate
whether ram-pressure is really responsible for the evolution of
galaxies in clusters. 
Most distant clusters observed in detail are rich ones, and their
temperatures are large ($k_{\rm B}T=6-10$ keV). Although the gas
density distributions of most of the clusters are not known, they are
assumed to be almost the same as those of nearby clusters. This is
supported by the fact that there is no evidence for the evolution of
luminosity-temperature relation among clusters at $z \sim 0.4$
(Mushotzky \& Scharf \markcite{ms1997}1997).  Thus, the models of
$k_{\rm B}T=8$ keV in \S\ref{sec:evol} can be applied to the
comparison. 

Using the population synthesis code made by Kodama \& Arimoto
\markcite{ka1997}(1997), the evolutions of color ($B-V$) and B-band 
luminosity in models A1 and B1 are calculated (Figure 5 and 6) on the
assumption that $\psi(t)=S_0$ for $-10<t<0$ Gyr. 
The evolutions in models A2 and B2 are similar to those in models A1
and B1, although the color and luminosity start to change later.
Model A1 shows the color and luminosity evolutions for a
infalling disk galaxy. 
In that model, ram-pressure sweeps the ISM away
at $t-t_{\rm cent}=-5\times 10^8$ yr, where $t_{\rm cent}$ is the
time when the galaxy reaches the cluster center ($t_{\rm cent} \approx
1.2$ Gyr). Before ram-pressure stripping occurs, the changes of color
and luminosity are very small. This indicates that the ram-pressure
compression does not induce observable star formation activity because 
the intrinsic scatter of color and luminosity among
galaxies are larger.
After the stripping occurs, the color of the model galaxy
rapidly becomes red and approaches the color of passively evolving
stellar system such as elliptical galaxies (Figure 5). Thus, the model
predicts that the fraction of blue disk galaxies is small in the
central part of rich clusters 
if significant fraction of the galaxies have
infallen from outside field.  

Recent {\em HST} observations indicate that there are proportionally more
spiral galaxies and fewer E/S0 galaxies in the high-redshift clusters
($z=0.3-0.5$) than in nearby clusters (Oemler et
al. \markcite{odb1997}1997 ; Dressler et al. \markcite{doc1997}1997 ;
Couch et al. \markcite{crs1998}1998) . Thus, some spiral galaxies may
have been transformed into E/S0 galaxies by $z=0$. This morphological
transformation may be explained by the models in \S\ref{sec:evol} as
follows. The luminosity of disk component of the model galaxy is
approximately proportional to that of model A1 as noted in
\S\ref{sec:evol}. Figure 6 shows the luminosity in model A1 becomes
40\% of the initial value after the model galaxy passes the cluster
center. On the other hand, the bulge of a disk galaxy generally has
little gas, and most of light from the bulge component is dominated by
passively evolving stellar system.  This means that the luminosity of
the bulge is not affected by ram-pressure very much. Even if the bulge
contains a large amount gas, the color and luminosity of the bulge
does not change significantly because in this case, the bulge
evolution follows model B1 (Figure 5 and 6). Therefore, the bulge to
disk luminosity ratio (B/D) changes when the galaxy radially infalls
into the cluster center.  For example if $\rm B/D = 0.2$ initially, it
increases to $\rm B/D = 0.5$ after the galaxy passes the cluster
center. This means Sb galaxies turn into S0 galaxies (Solanes et
al. \markcite{sss1989}1989). Therefore, the fraction of S0 galaxy
should be large in the central region of clusters. Moreover, the
number of S0 galaxy should increase in the all cluster as more field
spiral galaxies accrete to the cluster.

In summary, the ram-pressure from ICM indeed truncates star formation
activity in disk galaxies as Abraham et al. \markcite{ash1996}(1996)
speculate. Moreover, the above arguments show that if clusters have
been built up by the infall of blue field disk galaxies, the
ram-pressure model can explain the radial color and morphology
distribution observed in distant clusters, that is, the galaxy
population in the cluster changes gradually from a red, evolved,
early-type population in the inner part of the cluster to a
progressively blue, later-type population in the extensive outer
envelope of the cluster. The BOE can be explained by the high
proportion of the blue infalling galaxies. However, several
observational results may not be explained by our simple ram-pressure
model. In the rest of this section, a few examples are discussed.

First, Valluri \& Jog \markcite{vj1991}(1991) have shown that the
observed relationship between HI deficiency and size is opposite to
that expected from the ram-pressure model; they indicate that the
fraction of HI-deficient galaxies increases with optical size over
most of the size range.  On the other hand, other observations show
that HI-deficient galaxies are concentrated in the central part of
clusters (e.g. Cayatte et al. \markcite{cvb1990}1990). This is
consistent with the ram-pressure model. Valluri \& Jog
\markcite{vj1991}(1991) suggest that the disagreement results from
mass segregation in clusters. If the larger galaxies are more
concentrated toward the cluster core, they would be more severely
affected by the ICM than smaller galaxies.  Another solution may
exist. Since the luminosity of dwarf galaxies decreases after
ram-pressure truncates the star formation, the dwarf galaxies may
become too dark to be observed in cluster cores.

Second, Lea \& Henry \markcite{lh1988}(1988) report that the percentage
of blue objects in clusters seems to increase with the X-ray
luminosity for the most luminous clusters, and no correlation is
apparent at low luminosity. If ram-pressure is the only mechanism that
drives the evolution of galaxies in clusters, the fraction of blue
galaxies must always be high in low X-ray luminosity clusters, which
usually have low temperatures, because ram-pressure stripping is not
effective (see Model A4 in \S\ref{sec:evol}). 

Finally we would like to remark that although the effect of
ram-pressure is important
for cluster evolution, it cannot be denied
that other processes, such as galaxy harassment, affect some galaxies
in clusters.

\section{Conclusions}
\label{sec:conc}

We have quantitatively investigated the effect of ram-pressure on the
star formation rate of a radially infalling disk galaxy in a cluster
using a simple model of molecular cloud evolution. The primary results
of the study can be summarized as follows:

1. Since ram-pressure compresses the molecular gas, the star formation
rate of a disk galaxy increases to twice its initial value at
most as the galaxy approaches the center of a cluster with high gas
density and deep potential well, or with a central pressure of $\sim
10^{-2}\rm\: cm^{-3}\; keV$. 
However, this increase does not
affect the color of the galaxy significantly.

2. When the galaxy approaches closer to the cluster center 
($\lesssim 1$ Mpc from the center), the star formation rate of the
disk component rapidly drops due to ram-pressure stripping. As a
result, the galaxy turns red and the disk becomes dark. However, the
star formation rate of the bulge component does not significantly
change.

3. On the other hand, in a cluster with low 
gas density and shallow potential well, 
or the central pressure of $\sim 10^{-3}\rm\: cm^{-3}\; keV$, the
change of star formation rate is small. In these clusters neither
ram-pressure compression nor stripping is effective. 

4. The color and luminosity change induced by the ram-pressure effects
can explain the color and morphology distribution and the evolution of
galaxies observed in high-redshift clusters if the clusters have been
built up by accretion of field spiral galaxies.

\acknowledgments

We would like to thank I. Smail for his useful comments.  We are also
grateful to an anonymous referee for improving this paper. This work
was supported in part by the JSPS Research Fellowship for Young
Scientists.

\newpage

\include{table1}

\newpage

%\section*{Figure Captions}

\begin{figure}[htb]
 \label{f1a}
 \plotone{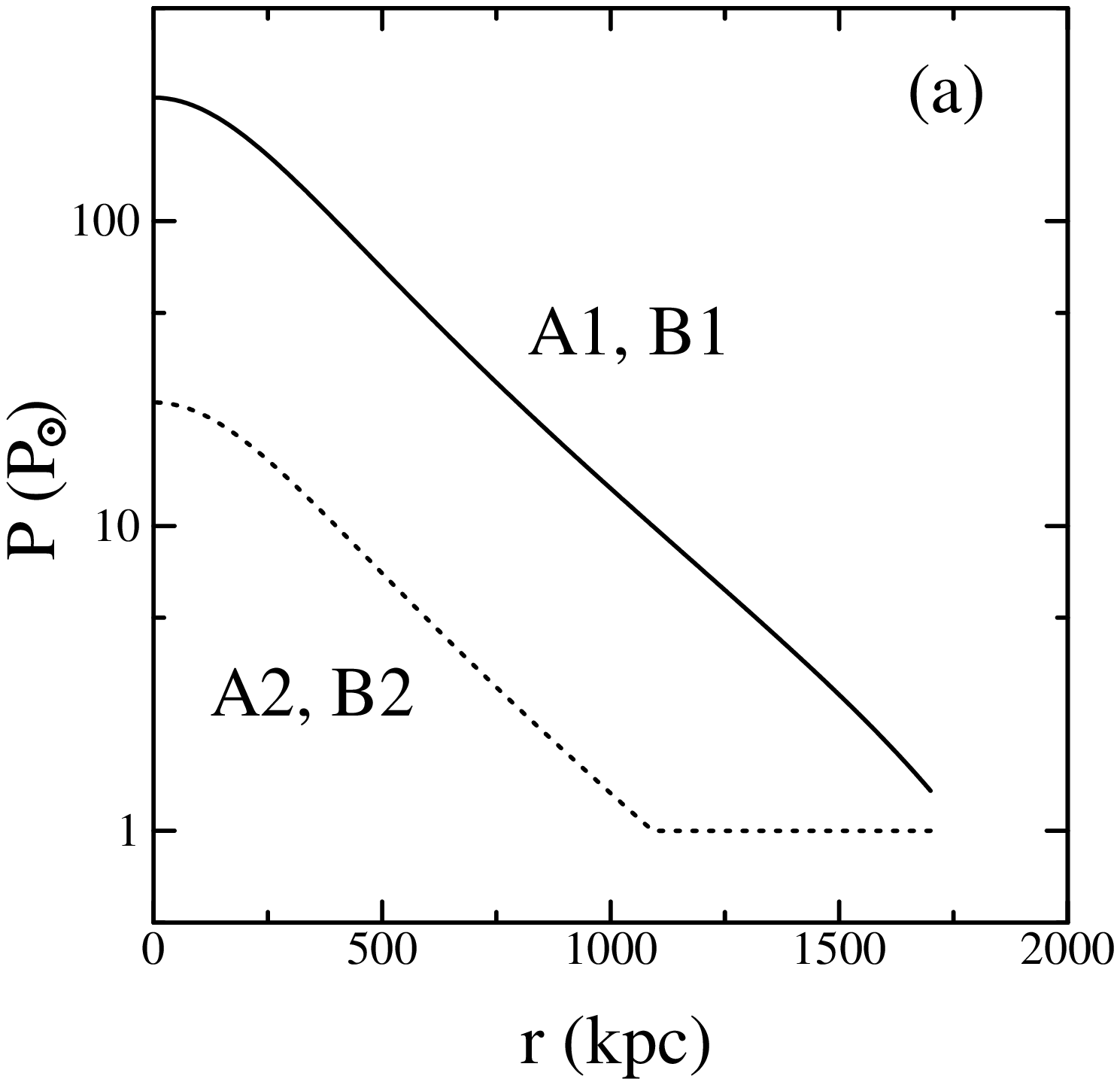}
\caption[f1a.eps]
{The pressure of a radially infalling galaxy as a function of
cluster radius. The units of pressure are $P_{\sun}=3\times 10^4\rm\;
cm^{-3}\; K$. (a) $k_{\rm B}T=8$ keV, (b) $k_{\rm B}T=2$ keV.
}
\end{figure}

\newpage

\begin{figure}[htb]
 \label{f1b}
 \plotone{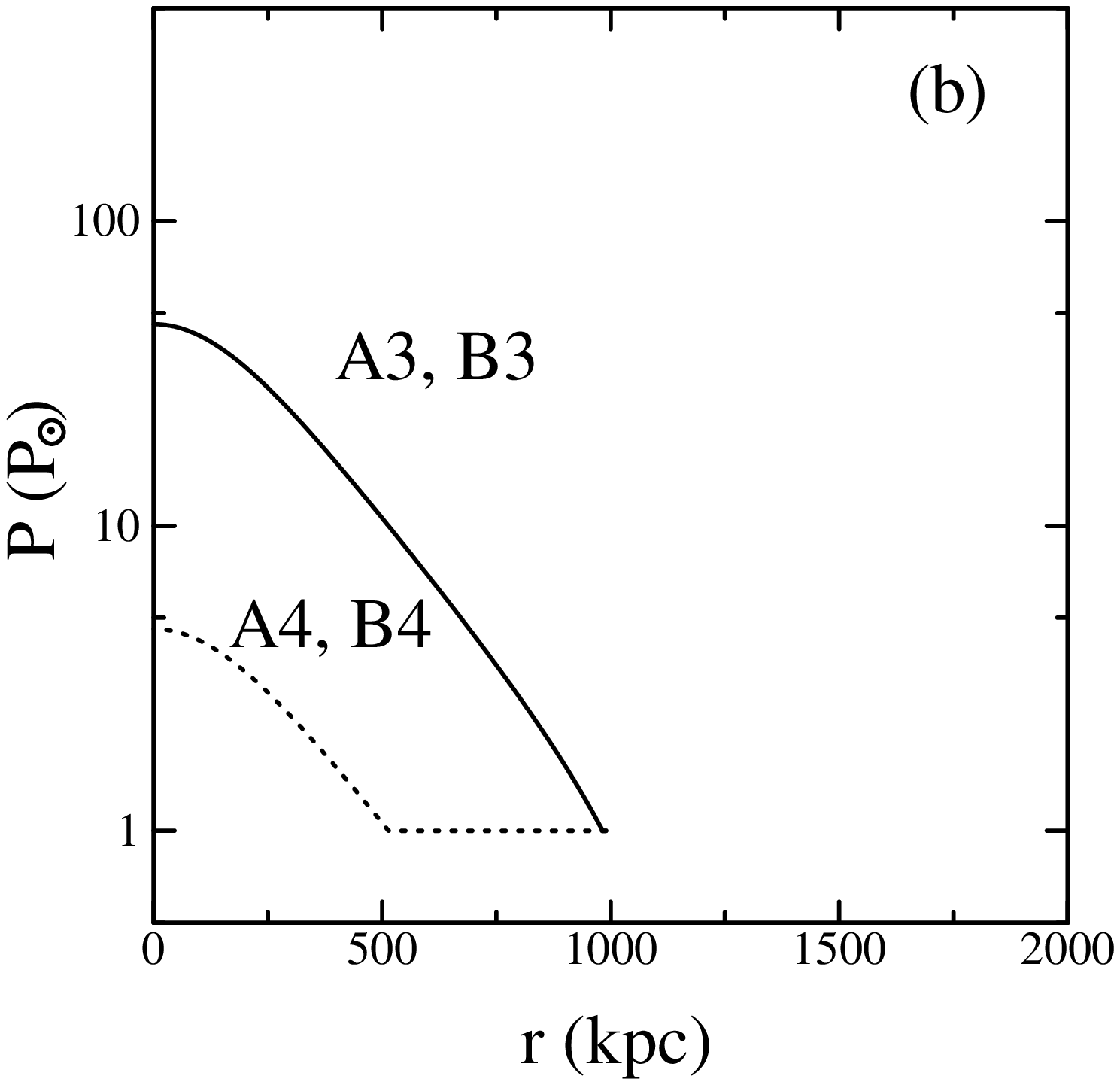}
\end{figure}

\newpage

\begin{figure}[htb]
 \label{f2a}
 \plotone{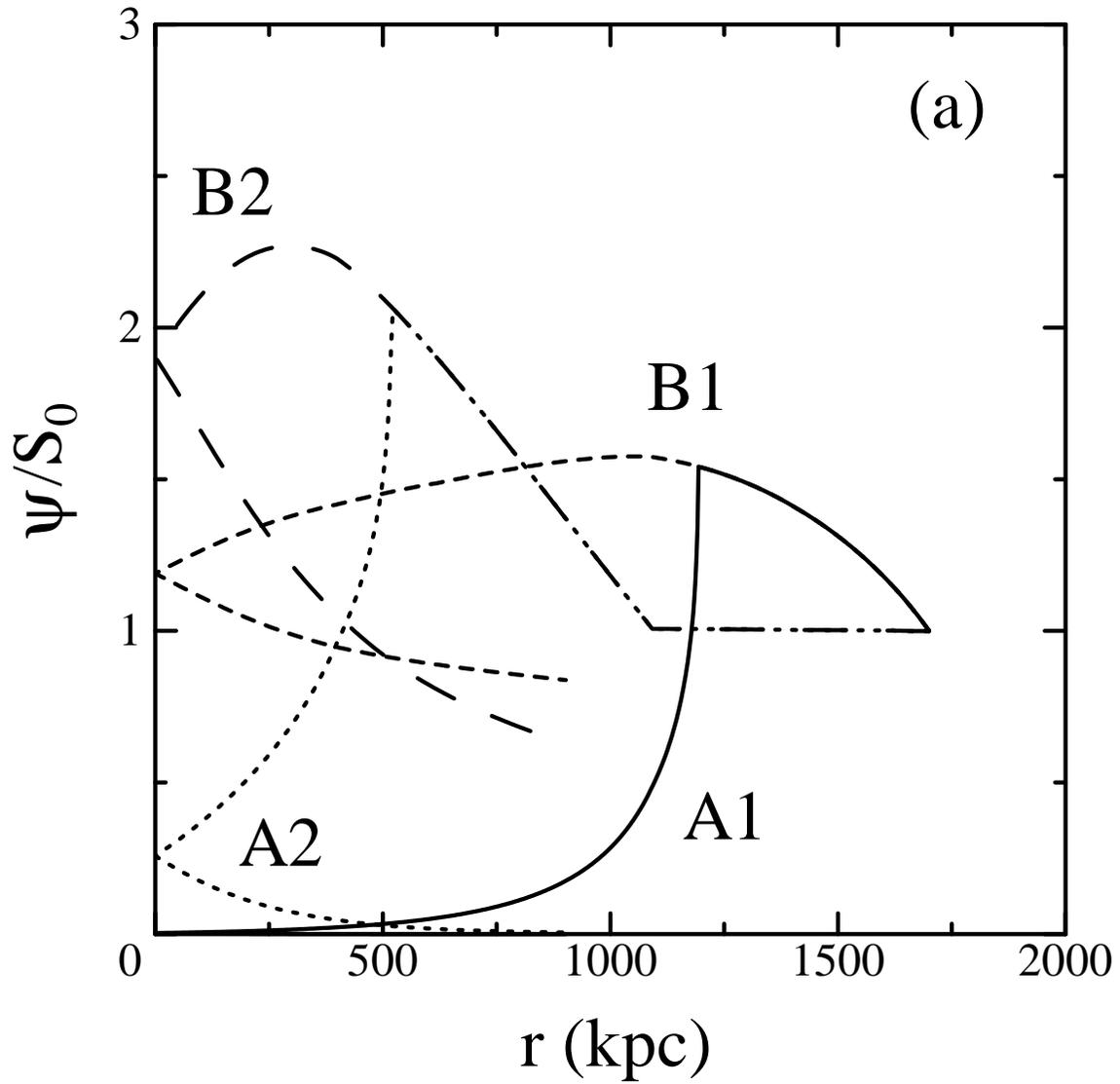}
\caption[f2a.eps]
{The normalized star formation rate 
of a radially infalling galaxy as a function of
cluster radius. (a) $k_{\rm B}T=8$ keV, (b) $k_{\rm B}T=2$ keV.
}
\end{figure}

\newpage 

\begin{figure}[htb]
 \label{f2b}
 \plotone{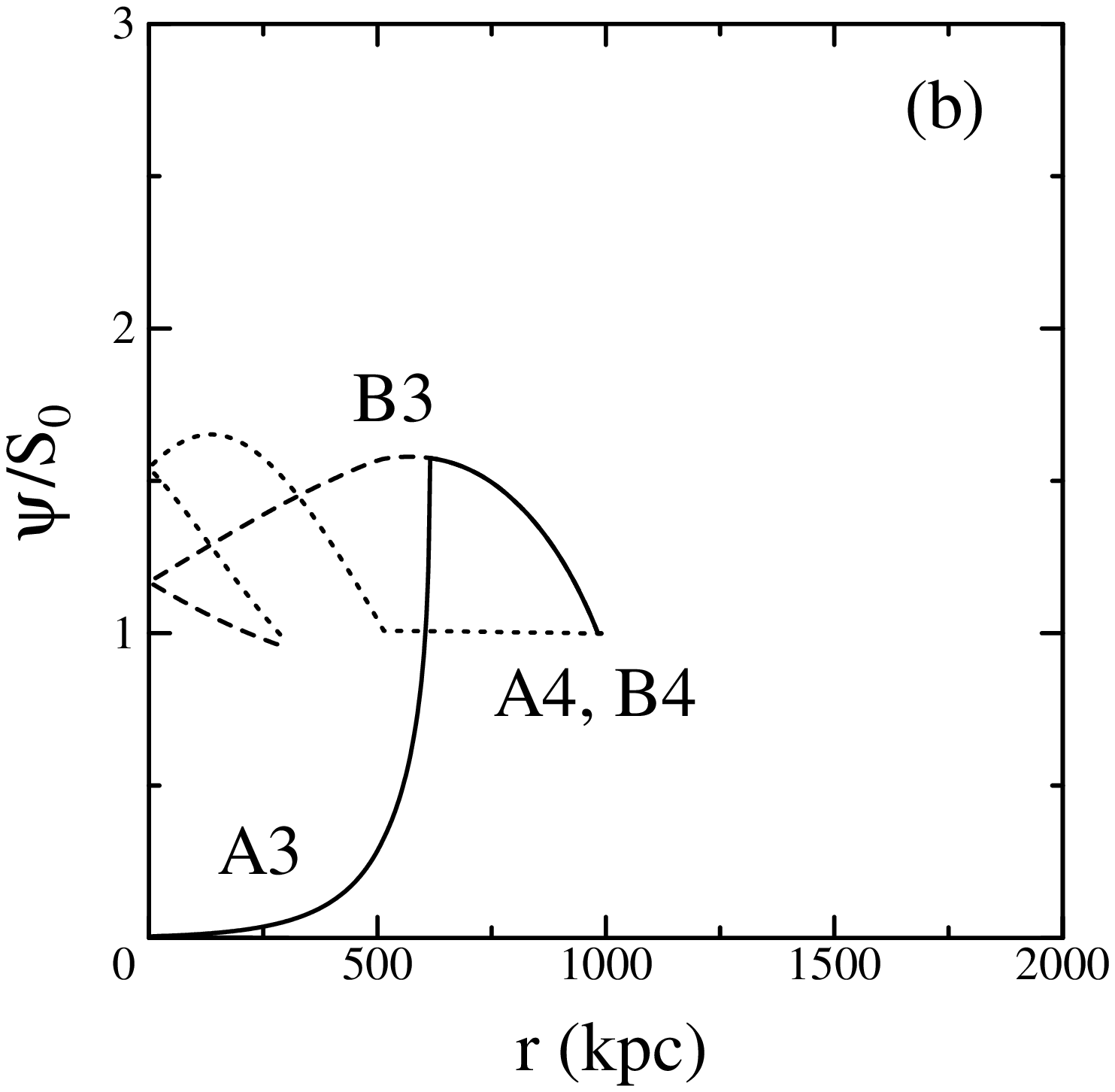}
\end{figure}

\newpage

\begin{figure}[htb]
 \label{f3a}
 \plotone{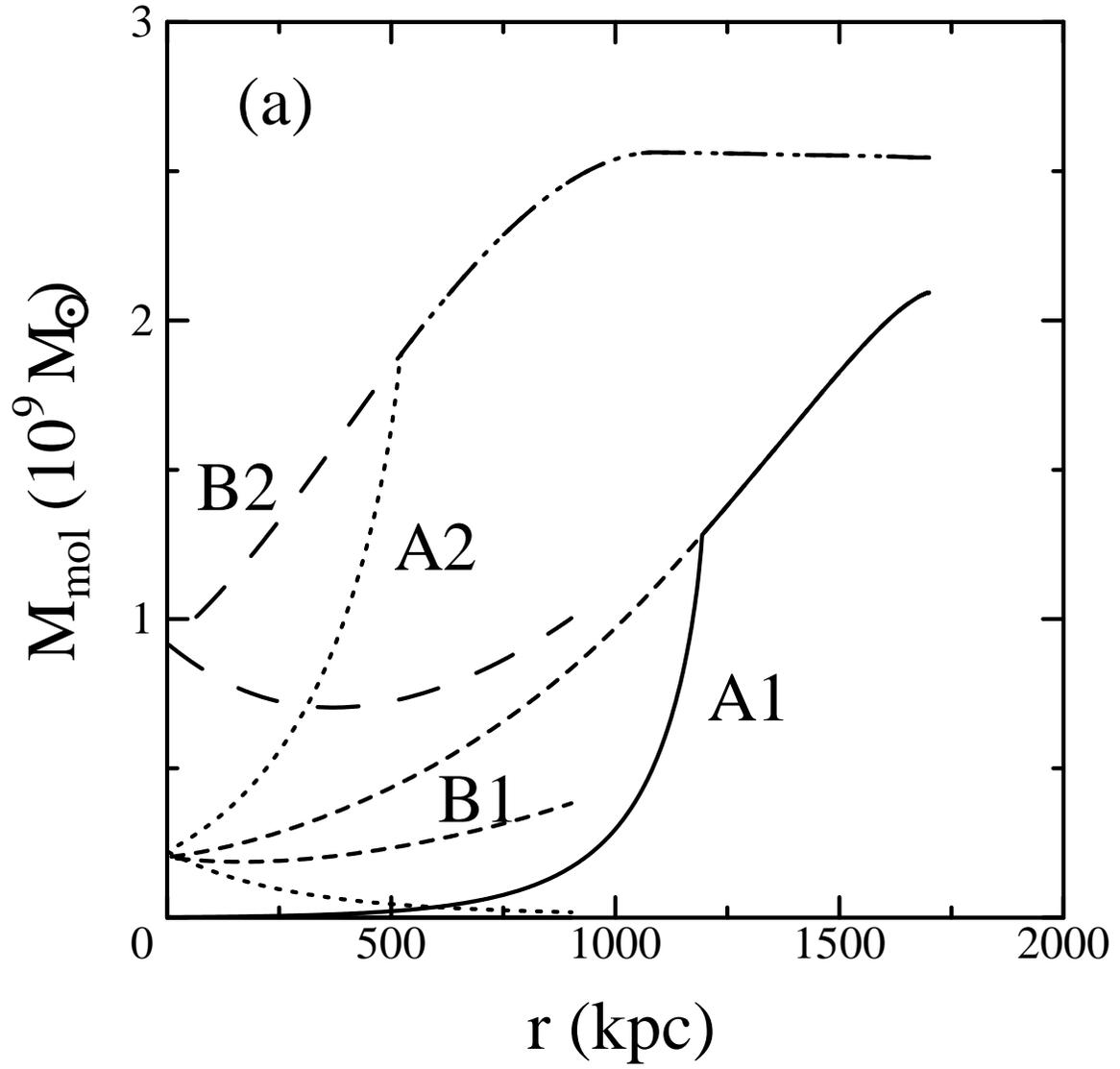}
\caption[f3a.eps]
{The mass of molecular gas of a radially 
infalling galaxy as a function of
cluster radius. (a) $k_{\rm B}T=8$ keV, (b) $k_{\rm B}T=2$ keV.
}
\end{figure}

\newpage

\begin{figure}[htb]
 \label{f3b}
 \plotone{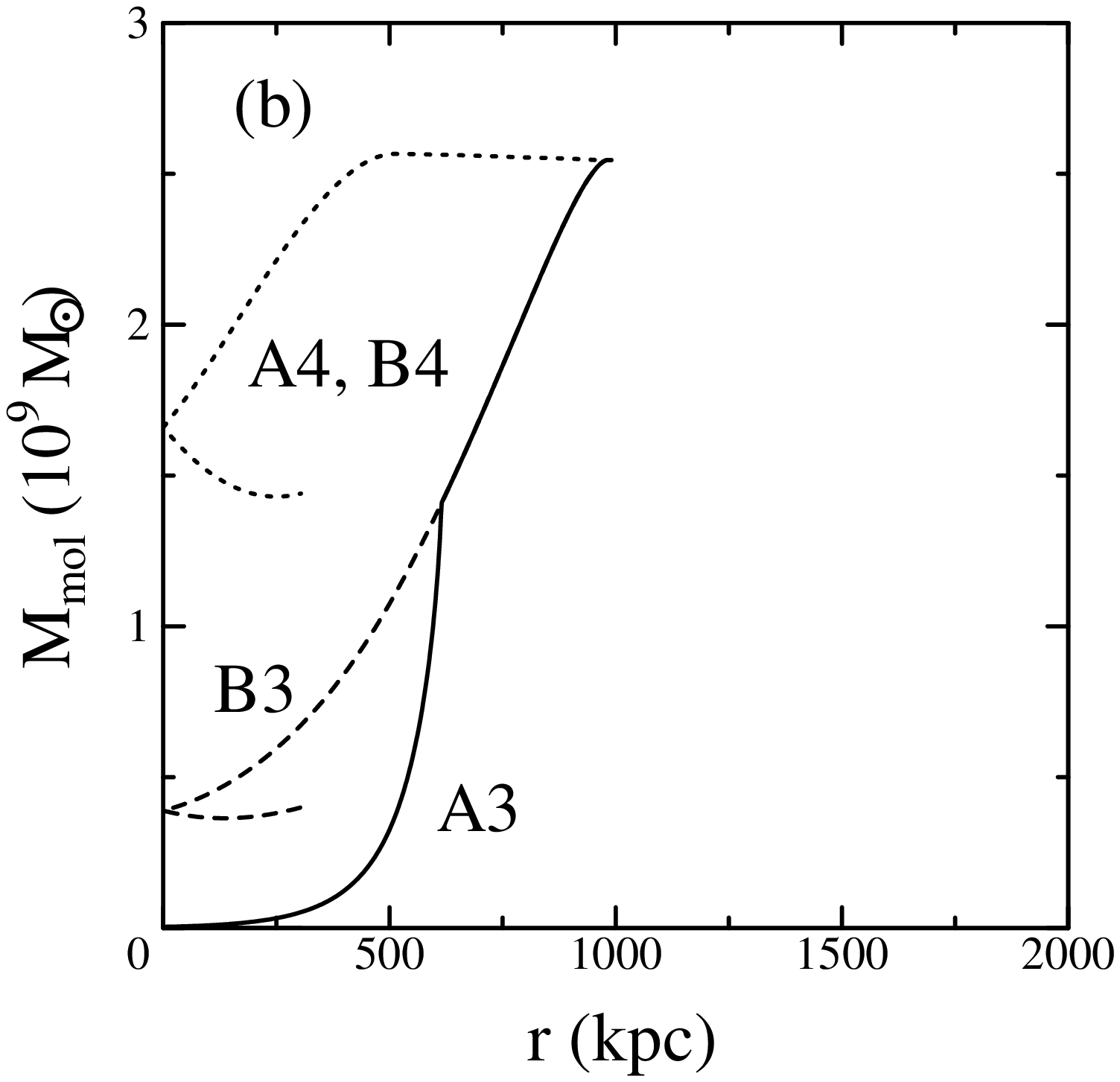}
\end{figure}

\newpage

\begin{figure}[htb]
 \label{f4}
 \plotone{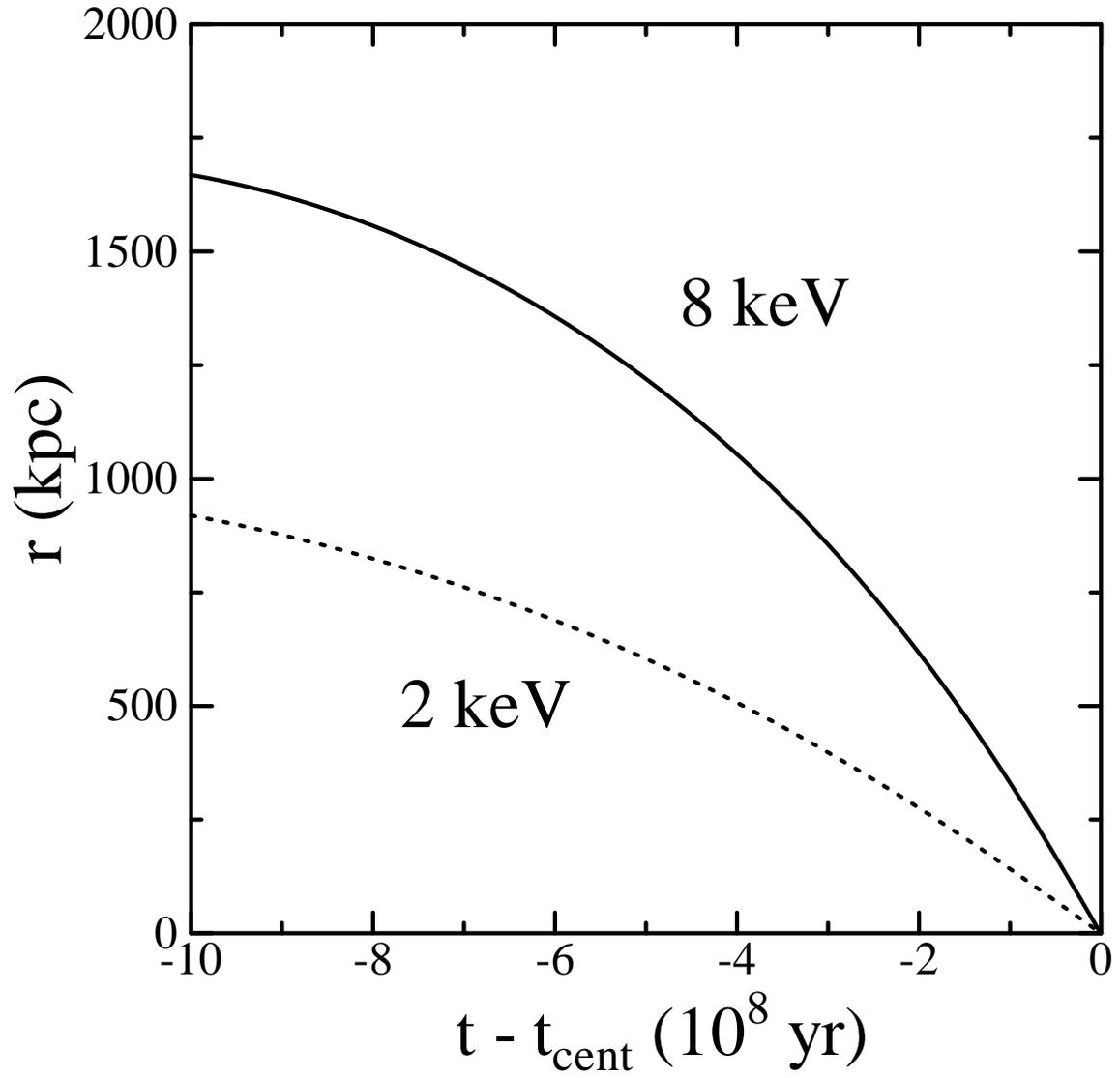}
\caption[f4.eps]
{The orbits of model galaxies. The time when the galaxies
reach the cluster center is referred as $t_{\rm cent}$.
}
\end{figure}

\newpage

\begin{figure}[htb]
 \label{f5}
 \plotone{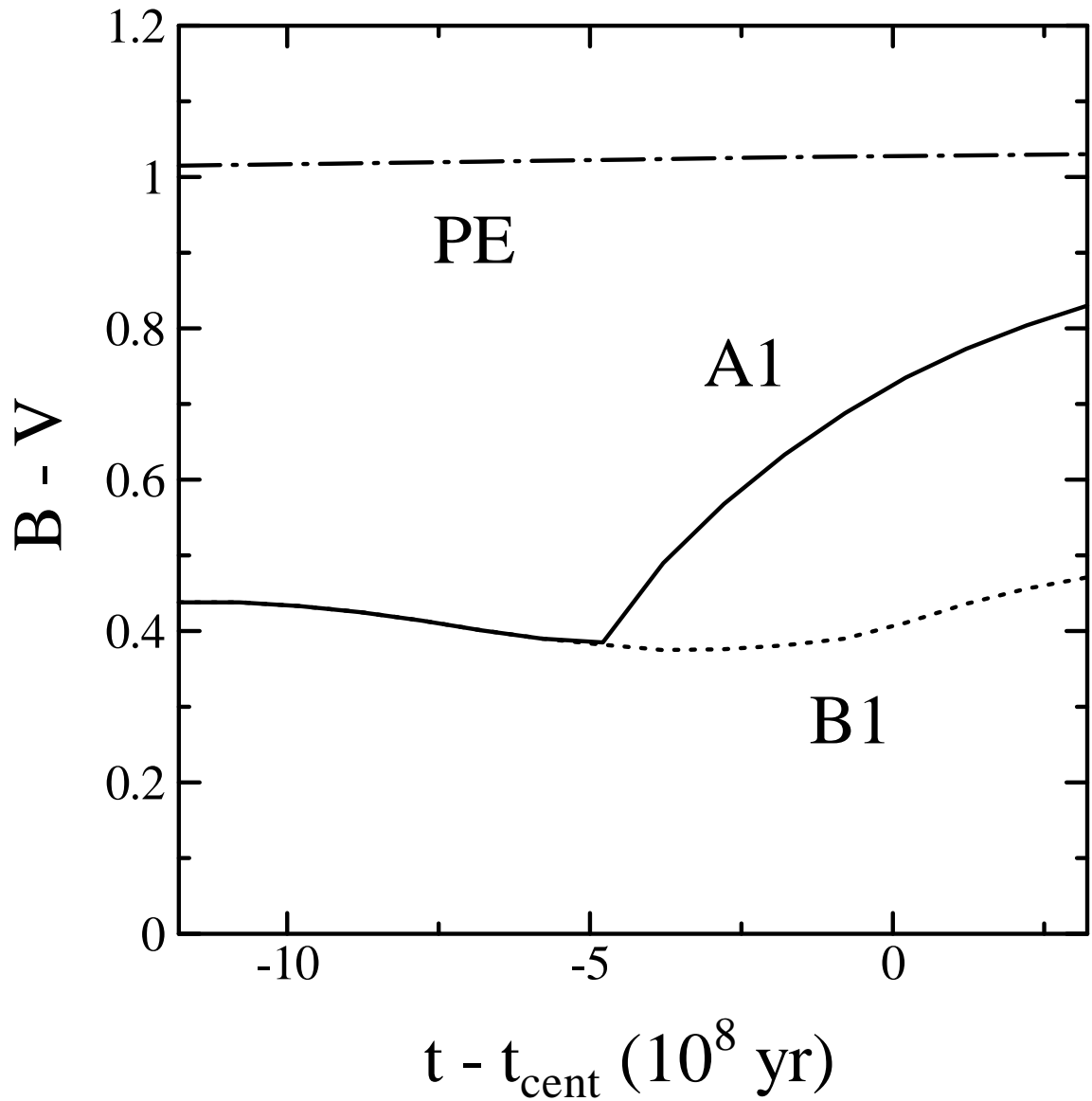}
\caption[f5.eps]
{The color evolution of model galaxies. The color of stellar
system which was simultaneously born at $t=-10$ Gyr with Salpeter
mass function and solar metal abundance is shown for comparison (PE).
}
\end{figure}

\newpage

\begin{figure}[htb]
 \label{f6}
 \plotone{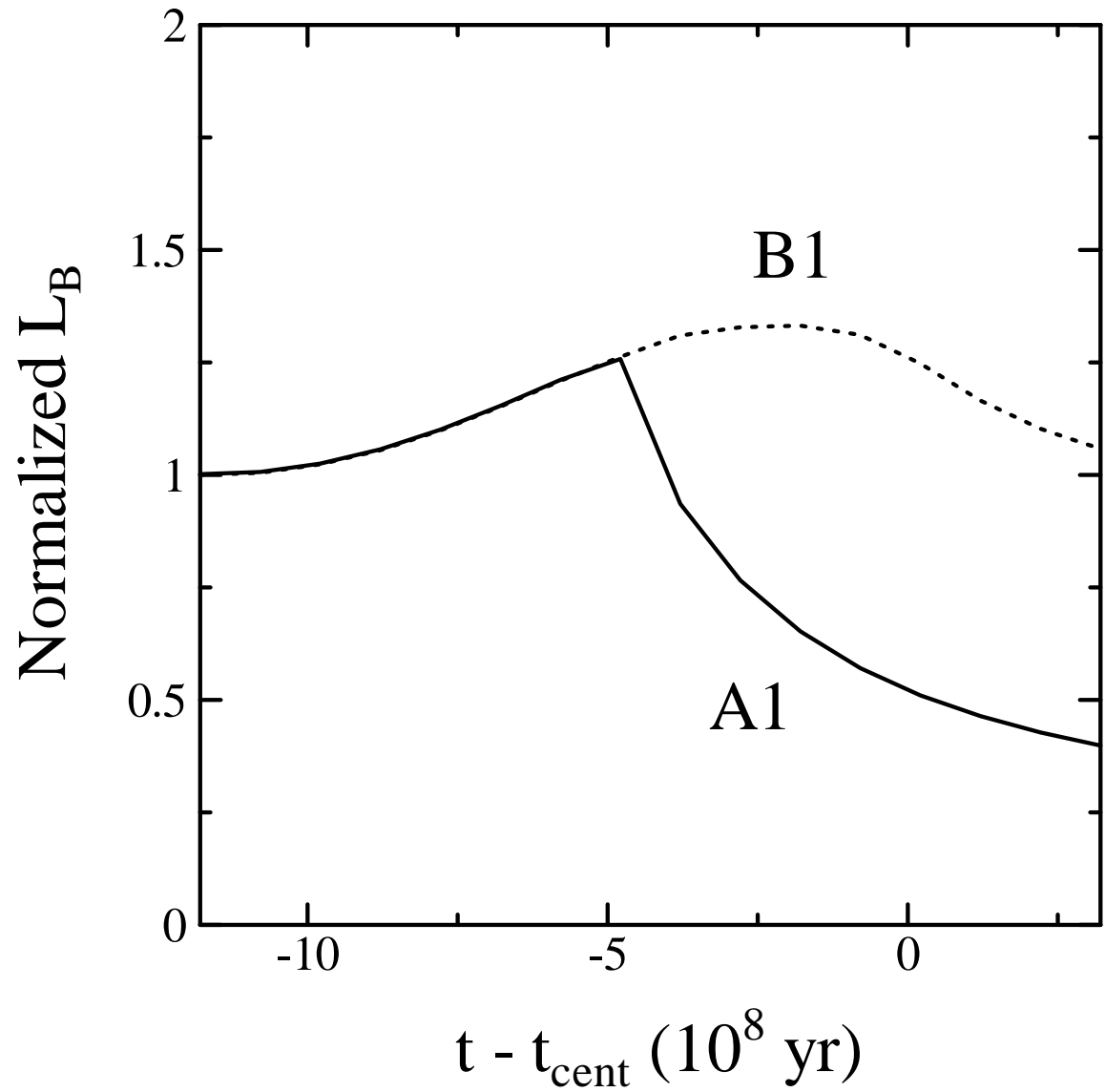}
\caption[f6.eps]
{The evolution of normalized B-band luminosity of model
galaxies.
}
\end{figure}

\end{document}

%% file: table1.tex
\begin{deluxetable}{cccccc}
\tablewidth{30pc}
\tablecaption{Models}
\tablehead{
\colhead{Model}               & \colhead{$k_{\rm B}T$} &
\colhead{$\rho_{\rm gas,0}$} & \colhead{$r_0$}    &
\colhead{$v_{\rm gal,0}$}    &    \colhead{Stripping} \\
\colhead{}  & \colhead{(keV)} &
\colhead{($10^{-28}\rm\; g\; cm^{-3}$)} & \colhead{(kpc)} &
\colhead{($\rm km\; s^{-1}$)}  & \colhead{}
}

\startdata
A1 &  8  &  83.5  & 1700 &   0 & yes \nl
A2 &  8  &  8.35  & 1700 &   0 & yes \nl
A3 &  2  &  83.5  & 990  & 129 & yes \nl
A4 &  2  &  8.35  & 990  & 129 & yes \nl
B1 &  8  &  83.5  & 1700 &   0 & no  \nl
B2 &  8  &  8.35  & 1700 &   0 & no  \nl
B3 &  2  &  83.5  & 990  & 129 & no  \nl
B4 &  2  &  8.35  & 990  & 129 & no  \nl
\enddata
\end{deluxetable}